# Comparison of the x-ray tube spectrum measurement using BGO, NaI, LYSO, and HPGe detectors in a preclinical mini-CT scanner: Monte Carlo simulation and practical experiment

Vahid Lohrabian[1], Alireza Kamali-Asl[1], Hossein Ghadiri Harvani[2], Seyed Rashid Hosseini Aghdam[1], Hossein Arabi[3]

[1] Department of Medical Radiation Engineering, University of Shahid Beheshti, Tehran, Iran
[2] Medical Physics and Biomedical Engineering Department, and Research Center for Molecular and Cellular Imaging, Tehran University of Medical Sciences
[3] Division of Nuclear Medicine and Molecular Imaging, Department of Medical Imaging, Geneva University Hospital, CH-1211 Geneva 4, Switzerland

**Abstract**

**Background:** In diagnostic x-ray computed tomography (CT) imaging, some applications, such as dose measurement using the Monte Carlo method and material decomposition using dual/multi-energy approaches, rely on accurate knowledge of the energy spectrum of the x-ray beam. In this regard, x-ray detectors providing an accurate estimation of the x-ray spectrum could greatly impact the quality of dual/multi-energy CT imaging and patient-specific dosimetry.

**Purpose:** The aim of this study is to estimate the intrinsic efficiency and energy resolution of different types of solid-state gamma-ray detectors in order to generate a precise dual-energy x-ray beam from the conventional x-ray tube using external x-ray filters.

**Materials and methods:** The x-ray spectrum of a clinical x-ray tube was experimentally measured using a high purity Germanium detector (HPGe) and the obtained spectrum validated by Monte Carlo (MC) simulations. The obtained x-ray spectrum from the experiment was employed to assess the energy resolution and detection efficiency of different inorganic scintillators and semiconductor-based solid-state detectors, namely HPGe, BGO, NaI, and LYSO, using MC simulations. The best performing detector was employed to experimentally create and measure a dual-energy x-ray spectrum through applying attenuating filters to the original x-ray beam.

**Results:** The simulation results indicated 9.16% energy resolution for the HPGe detector wherein the full width-at-half-maximum (FWHM) of the energy resolution for the HPGe detector was about $1/3^{rd}$ of the other inorganic detectors. The x-ray spectra estimated from the various source energies exhibited a good agreement between experimental and simulation results with a maximum difference of 6%. Owing to the high-energy discrimination power of the HPGe detector, a dual-energy x-ray spectrum was created and measured from the original x-ray spectrum using 0.5 and 4.5 mm Aluminum external filters, which involves 70 and 140 keV energy peaks with 8% overlap.

**Conclusion:** The experimental measurements and MC simulations of the HPGe detector exhibited close agreement in high-energy resolution estimation of the x-ray spectrum. Given the accurate measurement of the x-ray spectrum with the HPGe detector, a dual-energy x-ray spectrum was generated with minimal energy overlap using external x-ray filters.

# I. Introduction

Advances in detector technologies and accurate extraction of the energy spectrum of radioactive sources or x-ray generators have improved the quality and accuracy of x-ray imaging. In this regard, due to the low detection efficiency of gamma-ray detectors, the calibration of organic detection devices is still considered a significant challenge for gamma-rays with high energies. The reason for this is that the scintillation detectors do not generate any photopeak in their spectral response, especially for gamma rays with energies above 100 keV [1]. Precise knowledge about the x-ray energy spectrum of a diagnostic x-ray beam could aid to enhance the image quality and reliably estimate the absorbed dose to the patient. In preclinical imaging researches, varieties of x-ray spectra are employed depending on the subjects under study and the end-point of the tasks which may be far different from the ones used in clinical practices [2].

Sodium iodide activated with thallium, NaI(Tl), is the most widely used scintillation material. The NaI(Tl) and high pure germanium (HPGe) detectors, in addition to their applications in radiation and dosimetry systems, are considered ideal gamma-ray detectors owing to their good energy resolution and detection efficiency. HPGe detectors are made of extremely pure p-type or n-type germanium, wherein the active volume of these detectors is large enough for efficient and high-resolution gamma-ray spectroscopy. HPGe detectors have two advantages over germanium crystals were doped with lithium ions (GeLi) detectors: 1. they are able to function properly in the ambient temperature and 2. The production cost and procedure are much cheaper and easier. The resolution of an HPGe detector is usually expressed as the full width at half maximum (FWHM) for the 1332.5 keV energy peak of the Co-60 source. However, the energy resolution for a scintillation detector is usually expressed as the percentage of the relative efficiency for the energy peak of a Cs-137 source [3-5]. This parameter defines the capability of the detector to distinguish the different energy peaks in a multi-energy photon beam. In general, the energy resolution is commonly determined/affected by the factors such as; 1) fluctuation in the light generation for the same mono-energy irradiation, 2) the non-proportionality of light response, 3) statistical fluctuation in charge collection in the anode of the photomultiplier tube and 4) electronics noise [7, 8, 9]. In Monte Carlo simulation (particularly MCNP platform), Gaussian energy broadening (GEB) is a special order for tallies to better simulate a radiation detector whose energy peaks exhibit Gaussian energy broadening. Estimation of the Gaussian energy broadening parameters of the gamma-ray detectors plays a critical role in accurate modeling and interpretation of the resulting energy spectra [5-7].

In this regard, an energy broadening model for Bismuth germanium oxide (BGO) detector was developed by Askari et al. [8] wherein the GEB parameters for a 10 × 10 BGO array were calculated for the task of spectral response analysis in the presence of the wide energy range of gamma-rays in a simulation setting. Bugby et al. [9] conducted a comparison study between Thallium activated Cesium Iodide (CsI:Tl) and Gadolinium oxysulfide (GOS) scintillators for the use in a portable gamma camera using the GEB parameters of the Taheri et al. study [6]. This study demonstrated that the major difference between these two types of scintillator lies in their intrinsic spatial resolution, which was calculated to be 230 μm for the CsI:Tl scintillator and 1090 μm for the GOS scintillator. Goodle et al. [10] validated a Monte Carlo model for HPGe detector in order to predict the product ratios (chain) of activities in Fe, Cr, Ni, and stainless-steel foils. Ordonez et al. [11] investigated the full energy peak efficiency curves for an HPGe detector using MCNP6 and GEANT4. A Monte Carlo simulation of the efficiency of the HPGe detector was conducted by Azbouche et al. [12] to measure a large volume of environmental samples. The GEB parameters of an HPGe Detector were estimated by Eftekhari Zadeh et al. [13] to measure the energy resolution of GMX (GAMMA-X) series of coaxial detector systems. Moreover, the detection efficiency of a Compton suppression HPGe spectrometer was estimated in a work by Minh Tuan et al. [10], and the response function of a 3×3 inch NaI Scintillation detector for the energy range of 0.081 to 4.438 MeV was measured by Miri Hakimabad et al. [11].

Since the GEB parameters and detector energy resolution are the determining factors for accurate modeling of the detector response, the precise estimation of these parameters is the key for reliable evaluation of these detectors [11, 14, 15]. There have been always interests in designing detectors with the highest possible sensitivity (to detect very low activity concentration) and high spatial resolution (to localize the spatial distribution of the activity concentration). Normally, there is a fundamental tradeoff between improvements in sensitivity and spatial resolution of a detector, wherein strategies such as reducing the field of view (FOV) or modifying the distances between detector, object, and x-ray source (variable magnification) have been adopted to find an optimal compromise between these two key parameters [16-21].

HPGe is one of the highly accurate detectors for reliable measurement of the gamma-ray radionuclides and x-ray energy spectra [22-24]. Monte Carlo simulation is normally employed for precise modeling of the HPGe-based detection systems and estimation of the outcome spectra typically using either the Monte Carlo N-Particle radiation transport code (MCNP) or the Geant4 toolkit [25, 26]. Determination of the x-ray spectrum plays a key role in Monte Carlo-based dose calculations, dual- and/or multi-energy CT imaging, and material decomposition. Currently, there is a growing interest in patient-specific dose estimation from CT scans as well as patient-specific material/tissue decomposition using dual-or multi-energy CT imaging in clinical practices. Therefore, accurate x-ray spectroscopy approaches are highly required for such applications. However, due to the high photon flux, normally used in x-ray imaging, it is challenging to directly measure the spectrum of a CT scanner x-ray source. Different methods have been developed to estimate CT spectra which could be classified into two different groups, namely model-based (generating spectra from empirical or semi-empirical physical models), measurement-based (reconstructing/estimating spectra from measured data) [27].

One of the key components of CT scanners is the detection system which greatly contributes to the quality of the resulting images. Cadmium zinc telluride-based (CZT) semiconductor detectors are widely used in medical imaging systems owing to their excellent energy resolution, the capability of providing high spatial resolution, and proper detection efficiency for a wide range of gamma-ray energies [28, 29]. Other types of gamma-ray detectors introduced in CT scanners are inorganic scintillators crystals NaI (Tl) and Bismuth Germinate (BGO).

Today, dual- and/or multi-energy CT imaging systems have become a popular approach to discriminating biological tissues in clinical and research settings. In this framework, the image is formed by two different energies, wherein an algorithm of image processing is employed to extract/detect specific tissue types separately. The underlying idea behind this approach is to use two specific energy ranges with the maximum separation between their spectra. Hence, the accurate extraction of the x-ray tube spectra plays a key role in the overall performance of this approach [30-33].

In this research, we set out to measure the x-ray spectrum of a mini-CT scanner using several detector models in order to generate a dual-energy x-ray source through applying attenuating filters to the x-ray beam. To this end, Monte Carlo simulation was employed to evaluate different detector types and then the detector with the best/satisfactory performance was chosen for the practical experiment. The focuses will be on the evaluation of the key detection characteristics such as intrinsic efficiency and energy resolution. The comparison of the different detector types was carried out in Monte Carlo simulation as well as practical experiments.

## 2. Materials and Methods

The major goal of this study is to extract the energy spectrum of an x-ray tube in order to generate a dual-energy x-ray beam with two separates energy peaks. To this end, four different radiation detectors were examined in the MC simulation environment to choose/select the detector with the highest energy resolution and detection efficiency. The HPGe exhibited superior performance and hence was selected to experimentally measure the energy spectrum of the x-ray tube. To validate the experimentally obtained results, the x-ray tube was simulated in the MC simulation environment to create a baseline for comparison. Given the experimentally obtained x-ray spectrum, external attenuating filters were designed and applied to create a dual-energy x-ray beam with separate energy peaks of 70 and 140 keV. In the following, we elaborate on each of the above-mentioned steps to generate a dual-energy x-ray spectrum.

### 2. 1. Gamma-ray detectors

In order to detect and monitor ionizing radiation at higher energies, bulky scintillator materials were introduced in the form of a single crystal such as Tl-doped NaI and CsI single crystals [34]. One of the important parameters in the simulation of the scintillators with the MCNP4c (MCNP is an open-source

Monte Carlo N-Particle simulation code) is the Gaussian energy broadening (GEB) [6, 11]. The estimation of this parameter would aid precise simulation of the spectral response of the scintillator using the Monte Carlo framework. The focus of this study is on the inorganic as well as HPGe radiation detectors. They are several solid-state detectors such as Bismuth Germinate (BGO), sodium iodide (NaI), and Lutetium-yttrium oxyorthosilicate (LYSO) which are mostly used in SPECT, PET, CT, bone densitometers, and medical probes [35].

The x-ray spectrum of a mini-CT scanner for the energy range of 20keV to 160keV was measured in a laboratory setup using several scintillator gamma-ray detectors listed in Table 1 with their major specifications. Moreover, the energy spectrum of a portable cone-beam x-ray tube (Medex instrument, model GemX-160) with a focal spot area of 0.5×0.7 mm$^2$ (500 µm nominal focal spot) was estimated in the MC simulation environment (MCNPX platform). The x-ray tube with a LiPo battery power supply of 28-35 Vdc was employed which is capable of generating x-ray beams with energies ranging from 20 to 160 kVp (typically 30–140 kVp), with a maximum beam current of 2 mA (typically 0.1–2 mA). The spectrum of the x-ray tube was also estimated using the gamma- or x-ray detectors listed in table 1 using the Monte Carlo simulation. The HPGe detector exhibited the highest energy resolution compared to other semiconductors or scintillation detectors.

**Table 1.** Specification of the γ & x-ray crystals [9, 15, 35-37].

| Detector | BGO | LYSO | NaI | CWO | GOS |
|---|---|---|---|---|---|
| **Effective atomic number** | 75 | 66 | 56 | 65 | 61.1 |
| **Density (g/cm$^3$)** | 7.13 | 7.1 | 3.67 | 7.9 | 7.3 |
| **Peak emission (nm)** | 480 | 428 | 415 | 475 | 545 |
| **Chemical formula** | $Bi_4Ge_3O_{12}$ | $Lu_{2(1-x)}Y_{2x}SiO_5$ | NaI(Tl) | $CdWO_4$ | $Gd_2O_2S$ |

**2.2 Measurement of HPGe detector characteristics**

HPGe is the only radiation detection medium that provides sufficient information to accurately and reliably identify radionuclides from their mono-energy gamma-rays. HPGe detectors have almost 20 to 30 times more energy resolution than sodium iodide detectors. The detector used in this study was a GMX series HPGe coaxial detector system (GAMMA-X) which is a coaxial Germanium (Ge) detector with an ultra-thin entrance window. While most coaxial detectors have entrance windows from 500- to 1000-µm thick, the entrance window of the GAMMA-X detector is a 0.3-µm-thick and ion-implanted contact which extends the lower range of useful energies to around 3 keV. The characteristics of the HPGe detector are presented in Table 2.

**Table 2.** Characteristics of the HPGe detector.

| device type | model |
|---|---|
| detector model number | GMX40P4-83 |
| preamplifier model | A257N |
| cryostat configuration | CFG-PG4-1.2 |
| HV filter model | 138 EMI |

One of the limitations of the HPGe detectors is the loss of sensitivity or efficiency when they are exposed to high flux radiation. To address this issue in the experimental setting, a collimator is used to reduce the photon flux, which is normally made from lead or tungsten. The diameter of the hole created on the lead surface should meet two specific criteria: (1) the diameter aperture is sufficiently large to supply the detector with an acceptable flux of photon. (2) The diameter of the aperture is small enough that the photon flux does not saturate the detector. One of the practical issues when working with HPGe is that the radiation source should be in the form of a point source in order to avoid the early saturation of the HPGe detector. In this regard, a lead block with a thickness of 7cm and a hole with a diameter of 1.5 mm was placed against the x-ray tube beam to render a parallel and/or needle-shaped beam as illustrated in figure 1.

**Figure 1**

To obtain the final x-ray tube spectrum based on the signals acquired from the HPGe detector, certain corrections/modifications pertinent to the material and geometry of the detector should be made. These modifications include I. correction for the counts related to the characteristic X generated by the photoelectric interactions, II. Correction for the absolute efficiency of the detector depending on various factors such as the size, material, and radiation energy, and III. Correction for the background radiation. Figure 2 shows the escape fraction curve at different energies and the absolute efficiency of the HPGe detector obtained from the algorithm proposed in [13].

**Figure 2**

## 2.3 X-ray spectral analysis using IPEM78

A wide range of computational software has been developed for extraction and estimation of the x-ray spectrum through the selection of different filters as well as source and beam properties such as the half-value layer, energy, mAs, and exposure levels. In this study, we utilized the models of Birch and Marshall (IPEM Report 78) [38, 39] which employs an XCOM database to estimate linear attenuation coefficients for numerous materials as well as wide ranges of radiology and mammography x-ray spectra [40]. The spectra for tungsten target at the tube voltages varying from 30 kV to 150 kV and target angles from 6° to 22° were generated for this study. The entire spectra were generated at an energy step/bin of 0.5 kV.

**2.4 X-ray spectral analysis using Spektr**

The Spektre toolset, based on the TASMIP algorithm of Boone and Seibert [41], provides realistic x-ray spectra across a wide range of kVp and beam filtration, wherein a databank for atomic number elements (Z = 1-92) and combinations of materials from the National Institute of Standards and Technology (NIST) is exploited. This toolset was used to generate x-ray beams in the energy range of 1-150 kV with 1 kV energy steps.

**2.5. Monte Carlo simulations**

The modeling of various inorganic scintillators and semiconductor detectors was performed using the MCNP software. There are three different methods for evaluation of the detector efficiency, namely the semi-empirical, Monte Carlo simulation, and the direct mathematical methods. In this study, the Monte Carlo simulation was employed to evaluate the response of the different detectors wherein the same simulation settings such as positioning, distances, and sources were utilized. Gaussian Energy Broadening (GEB) of all radiation detectors according to Table 3, were incorporated in the MC simulation [6, 8, 11, 13, 42]. The detector efficiency was determined using F8 tally, which is specific for pulse height analysis without any variance reduction. The geometry sought to simulate the detectors was a cylinder with a size of 7.62cm (height) × 7.62cm (diameter). The absolute detector efficiency was calculated by the pulse-height distribution per photon emitted from an x-ray source with kVp = 140 (typical CT scan x-ray spectrum). The statistical uncertainty was less than 1% in the entire MC simulations. The energy cutoff was set to 0.5 and 0.01 MeV for electrons and photons, respectively. The number of histories in each MC simulation was equal to $5\times10^8$. Figure 3 illustrates the schematic sketch of the simulated HPGe detector.

**Figure 3**

**2.6. Detector efficiency**

Detection efficiency is one of the key parameters in spectrometry and x-ray imaging systems which is specified by the intrinsic and geometry efficiency. Intrinsic efficiency is the probability that an incident photon in the detector will produce a meaningful pulse/signal. Geometry efficiency is the fraction of arrived photons on the detector surface per number of emitted photons from a source. The product of the intrinsic and geometry efficiencies gives absolute efficiency which depends on photon energy and source to detector distance. The detector efficiency can be determined via the count rate measurement of a calibrated source (with kwon energy spectrum) or Monte Carlo (MC) simulations.

Normally the entire emitted radiations wound not arrive at the detector's surface and all of the incident radiations would not interact in the sensitive volume depending on the detector sensitivity and distance from the source. Detector efficiency is divided into two intrinsic and geometric components. Detector efficiency usually depends on material density, sensitive volume size, radiation type, radiation energy, and connected electronic systems [22, 24, 43].

The intrinsic efficiency (Eq. 1) is defined as the number of recorded photons in the detector per number of incident photons to the surface of the detector.

$$\varepsilon_{int} = \frac{number\ of\ recorded\ on\ detector}{number\ of\ incident\ on\ detector} \qquad (1)$$

The geometry efficiency ($\varepsilon_g$) only depends on the detector size and the distance from the source. This also depends on the solid angle of the detector seen from the actual source position which is obtained from Eq. 2 [44]:

$$\varepsilon_g = \frac{1}{2}(1 - \frac{d}{\sqrt{d^2 + R^2}}) \qquad (2)$$

Where $d$ is the distance between source and detector and $R$ is the detector radius. If the number of recorded events on the active volume is calculated per number of photons emitted by the source, a new parameter could be defined as absolute efficiencies ($\varepsilon_{abs}$) formulated in Eq. 3 [44].

$$\varepsilon_{abs} = \varepsilon_{int} \times \varepsilon_g \qquad (3)$$

**2.7. Energy resolution**

Radiation spectrometry can also be examined by recording the response to a monoenergetic source of radiation. This procedure can be repeated for different levels of energy. In this regard, the energy resolution is defined as the ability of the detector to accurately determine the energy of the incident radiation [44]. Solid-state semiconductor detectors can have an energy resolution of less than 1%, while scintillation detectors used in gamma-ray spectroscopy normally exhibit an energy resolution of 3–10% [45, 46].

The response function of a detector in a laboratory setting is normally expressed in the form of a Gaussian function. The Gaussian energy broadening, which is the key parameter to accurately simulate and assess the different detector performance, is formulated in Eq. 4.

$$f(E) = Ce^{-(\frac{E-E_0}{A})^2} \qquad (4)$$

Here, $E$ is the broadened energy, $E_0$ is the energy of the source, C is a normalization constant, and $A$ is a factor related to the full width at half maximum (FWHM) of the Gaussian function calculated by Eq. 5.

$$A = \frac{FWHM}{2\sqrt{\ln 2}} \quad (5)$$

The FWHM in Eq. 5 is calculated from the Eq. 6 wherein a, b, and c are the specific parameters of the GEB, and $E$ is the energy recorded in the detector. These parameters are shown in Table 3 for some well-known detector examples.

$$FWHM = a + b\sqrt{E + cE^2} \quad (6)$$

The energy resolution of the detector is defined as the FWHM divided by the photo-peak centroid $H_{max}$. The energy resolution ($R$) is thus calculated from Eq. 7

$$R\% = \frac{FWHM}{H_{max}} \quad (7)$$

Table 3. Gaussian energy broadening of some radiation detector.

| Crystal | Energy range (MeV) | Fitting parameter (a) | Fitting parameter (b) | Fitting parameter (c) |
|---|---|---|---|---|
| BGO | 0.059 keV-1.33 MeV | 1.56E-03 | 1.216E-01 | 2.455 |
| LYSO | 20 keV- 1.43 MeV | -6.387E-03 | 1.097E-01 | 6.574E-01 |
| NaI | 0.062keV-1.35MeV | -2.4E-03 | 5.165E-02 | 2.858 |
| HPGE | 40 keV-1.46MeV | 5.868E-04 | 3.951E-04 | 7.467 |

## 2.8. Dual-energy CT

Dual-energy CT (DECT) is one of the commonly used techniques in the medicine and industry for the separation/decomposition of the underlying elements of the material. The primary goal of the DECT scan is the decomposition and/or discrimination of the different elements or biological tissues. To this end, it is necessary to create two spectra with different energies that are able to render maximum differentiation between the different materials or tissues. In DECT imaging, two images are obtained with two different x-ray spectra, which can be generated by modifying the x-ray tube voltages or altering the x-ray spectrum using external filters. In this research, different spectra have been created by applying external filters to the x-ray beam.

The x-ray source (GMX-160) used in this experiment was able to produce x-ray spectra ranging from 20 kVp to 160 kVp. To create the maximum separation between the two x-ray spectra for the purpose of dual-CT imaging, aluminum filters with optimal thicknesses of 0.5 to 6 mm were employed to attenuate the original x-ray beam. These filters (thicknesses of the aluminum filter) have been optimized through Monte Carlo simulation using the NIST (National Institute of Standard and Technology) library.

## 3. Results

In order to record the spectrum of the x-ray with an HPGe detector, the different channels connected to the detector must first be calibrated with known energies. Regarding the energy range generated by the x-ray source, an Americium-241 source was employed with γ-ray energies of 59.5 (35.9%) keV for the primary radiation and 13.9keV (13%), 17.6keV (20.2%), and 26.4 keV (5.2%) for the secondary's. The distance between the source and the detector was set about 2 meters to avoid saturation of the detector due to high photon flux and maintaining good statistics at the same time. After the calibration of the HPGe energy channels, the x-ray tube was set at an energy of 140 kVp, a current of 0.1 mA, and an exposure time of 10 seconds. The spectrum of the resulted x-ray beam was recorded by the calibrated HPGe detector. The calculated pulse height distribution (the spectrum of the tube at specific energy) for a 140 kVp x-ray source is depicted in figure 4.

**Figure 4**

The Monte Carlo simulation results exhibited superior energy resolution of the HPGe detector over the other detectors as shown in Table 4. Thus, the HPGe detector was selected to experimentally measure the spectrum of the x-ray tube. To validate the spectra obtained from the practical experiment, the measured GEB parameters were utilized in the Monte Carlo simulation (using the MCNP4c package). As Figure 5 shows, the HPGe detector resulted in remarkably higher energy resolution compared to the other detectors. In figure 4 two energy peaks of 57.6 and 69.2 are clearly visible/distinguishable in the energy spectrum obtained from the HPGe detector. The simulation results confirmed the experimental measurements where the maximum difference of 5% in energy resolution was observed for the HPGe detector.

**Figure 5**

Table 4 summarizes the energy resolution and FWHM obtained from the simulation of the different detectors. It should be mentioned that the reported results are corresponding to the pulse height distribution of the detector response function for 140 kVp x-ray energy calculated by Eq. 8 ($H_{max}$ is the pulse height distribution).

$$Resolution = \frac{FWHM}{H_{max}} \times 100 \tag{8}$$

**Table 4.** FWHM and energy resolution for the different detectors measured for an x-ray beam with energy of 140 kVp.

| Detector | FWHM (%) | $H_{max}$ | Resolution (%) |
|---|---|---|---|
| HPGe | 0.5 | 0.060 | 9.16 |
| NaI | 1.4 | 0.060 | 23.33 |
| LYSO | 2.2 | 0.060 | 37.55 |
| BGO | 3.7 | 0.060 | 61.66 |

In the field of computed tomography imaging, one of the important applications of x-ray spectroscopy is to perform dosimetry using the Monte Carlo method as well as the analysis of the materials by dual-energy and multi-energy scanning approaches. In this study, HPGe detector was employed to carry out the x-ray spectroscopy experimentally as well as using Monte Carlo, IPEM78, and Spektr simulation platforms. Figure 6 illustrates the comparison between the spectra measured experimentally by the HPGe detector and by the Monte Carlo, IPEM78, and Spektr modeling softwares for x-ray energies of 60, 80, 100, and 140 KVp.

**Figure 6.**

For the task of dual-energy CT scanning, the key issue is to generate two x-ray spectra with a minimum overlap which leads to maximal discrimination of the underlying component of the subjects under study. Figure 7 shows the generated dual-energy x-ray spectrum obtained from applying the external filters (aluminum filters with thicknesses of 0.5 to 4.5 mm) to the original x-ray beam with an energy of 140 Kvp. This spectrum has been measured by the HPGe detector which exhibits two separable peaks with minimal tails overlap.

**Figure 7**

## 4. Discussion

The Monte Carlo method is a powerful and precise tool for the calculation and analysis of the x- and gamma-ray spectra. These spectra would be used for the tasks of dosimetry, image formation/correction, and material decomposition using dual- or multi-energy CT scanning. This study focused on MC simulation and comparison of the several detectors for the energy ranges commonly used in clinical CT scanning. The pulse high distribution increased when the energy of the x-ray beam increased for the entire detectors up to the energy of 60 keV and then dropped to zero at 140 keV energy. The reason for this observation might be due to the incorporation of the Gaussian Energy Broadening (GEB) into the MC simulation. Regarding the characteristics of the detectors such as density, effective atomic number, and different GEB parameters, there was a particular pulse high spectrum for each detector. However, there would be the same pulse high spectra for all detectors, if we do not consider the GEB parameters in the MC simulation.

Crystal characteristics, impurity, traps, degradation due to long time usage, and instability of the temperature could be the contributory factors to determine the energy resolution of a detector quantified by FWHM. The results suggest that high purity germanium crystal (HPGe) has a narrow energy FWHM compared to the other crystals measured in the same situations, owing to the physical characteristics of the HPGe detector which is made of semiconductor materials. These types of detectors allow for fast, high energy, and spatial resolution image acquisition due to the small and efficient detection medium. On the other hand, the MC simulation results showed a poor energy resolution (large FWHM) for the Bismuth Germinate (BGO) detector, though this detector has a high density as well as the atomic number which render this detector highly sensitive to the incident radiation. The low light gain of the BGO detector is its major disadvantage which results in poor energy discrimination power. Nevertheless, this detector is widely used in nuclear medicine imaging and dosimetry. The Sodom Iodide (NaI) crystal has relatively good energy resolution (exhibiting relatively small energy FWHM) when using a 140 keV x-ray beam. The NaI crystal is transparent with a relatively high light gain which makes it a scintillator detector with good energy resolution. However, NaI crystal is not suitable for CT scanner because this crystal should be coupled with PMTs.

The simulated x-ray spectra were compared with spectra generated by Spektr, IPEM78, and practical measurement at the same kVp. The experimental energy spectrum measurement exhibited good agreement with the Monte Carlo simulation results which demonstrates the superior capability of the HPGe detector to resolve the significant properties of the x-ray energy spectrum. Figure 8 shows the comparison between the x-ray spectrum obtained from different experimental measurements and simulation models including the simulation and experimental estimation conducted by Ay et al [39], Expectation-Maximization modeling by Duan et al [27], PENELOPE simulation by Jia et. al [2], practical measurement by Primak et al [47], the model developed by Punnoose et al [41] based on Spektr version 3.0 platform. The graphs in figure 8 exhibited good agreement between the x-ray spectra observed in this study and the different analytical models and experimental measurements reported in the literature. The small differences in these graphs were due to the different types of filters used for the x-ray beam and the different detector materials.

## 5. Conclusion

In the present work, Monte Carlo simulations of some commonly used detectors, namely HPGe, BGO, LYSO, and NaI were carried out to calculate the detector efficiency and energy resolution of these detectors for the photon energy ranging from 20 keV to 140 keV. Given the superior energy resolution of the HPGe detector, this detector was selected to practically measure the energy spectrum of a mini-CT scanner in order to generate a dual-energy X-ray beam. The practical spectrum obtained from the HPGe detector was in close agreement with the Monte Carlo simulation. External attenuating filters (0.5 and 4.5 mm Al) were applied to the x-ray beam which resulted in a dual-energy x-ray spectrum with minimal overlap and two distinct energy peaks.

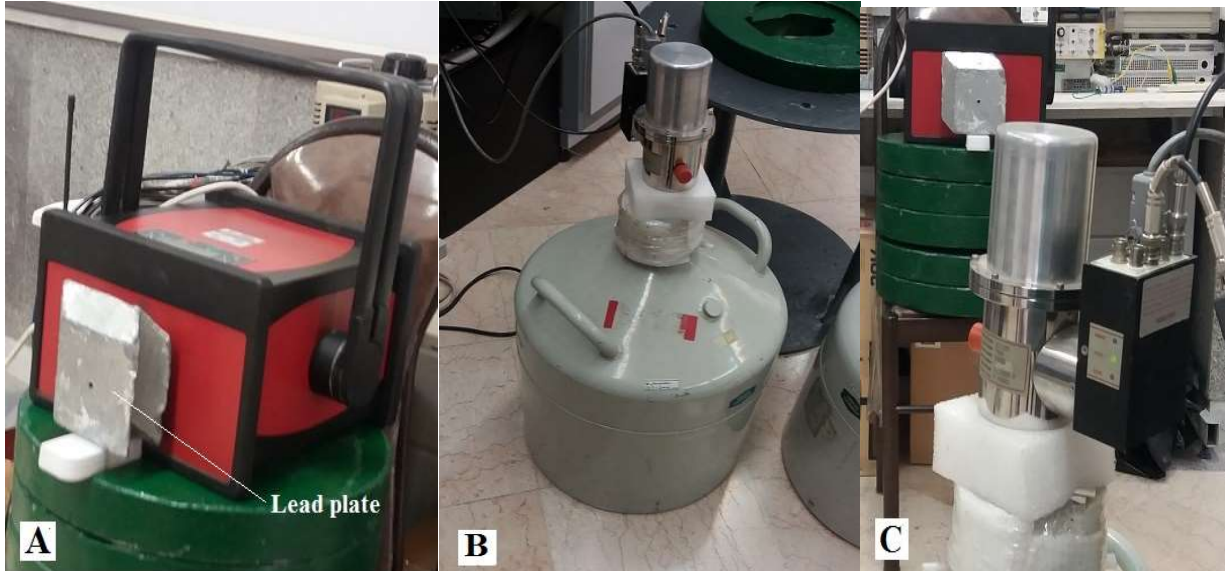

**Figure1.** A) The x-ray tube and the lead block to render a point source. B) HPGe coaxial detector model number GMX40P4-83. C) X-ray tube, lead block, and HPGe detector under test condition.

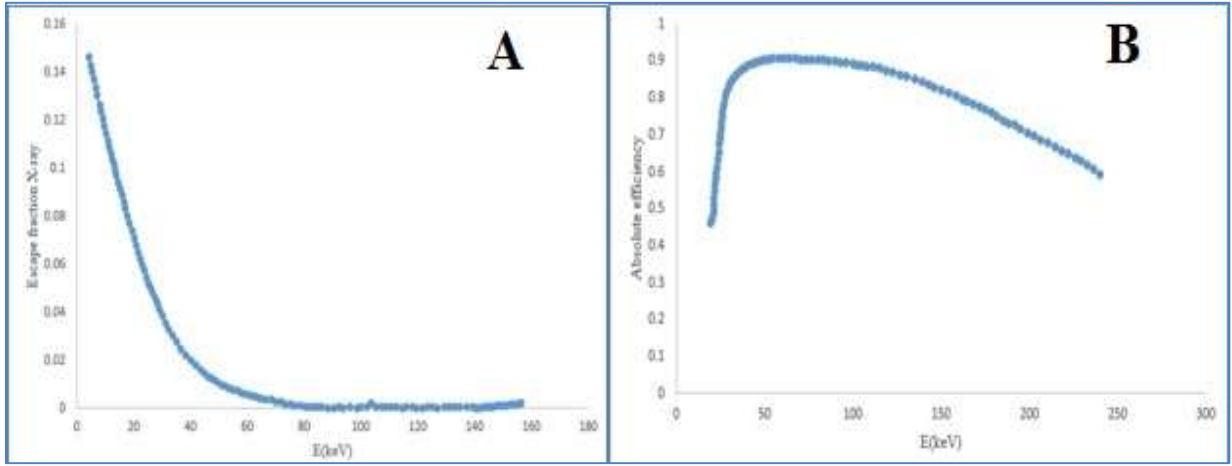

**Figure 2.** A) Absolute efficiency curves as a function of photon energy and B) the escape fraction of x-rays.

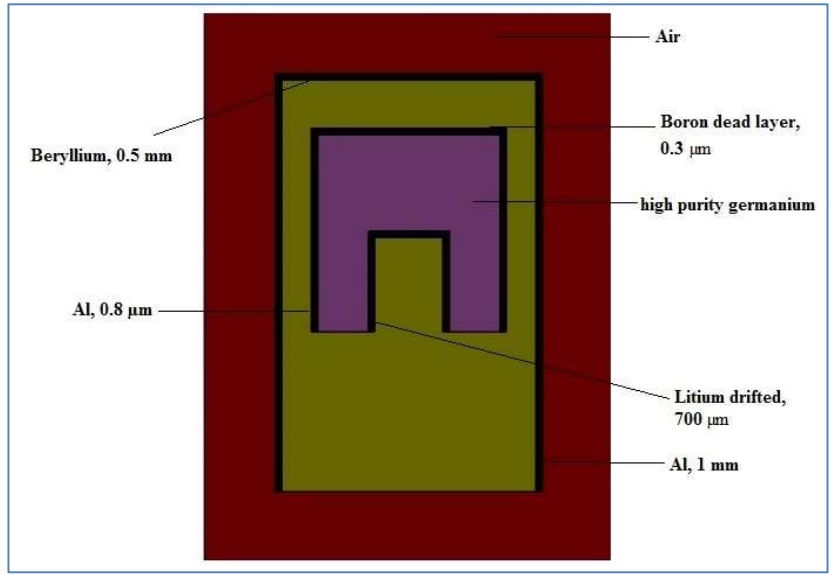

**Figure 3.** Schematic geometry of the simulated HPGe detector.

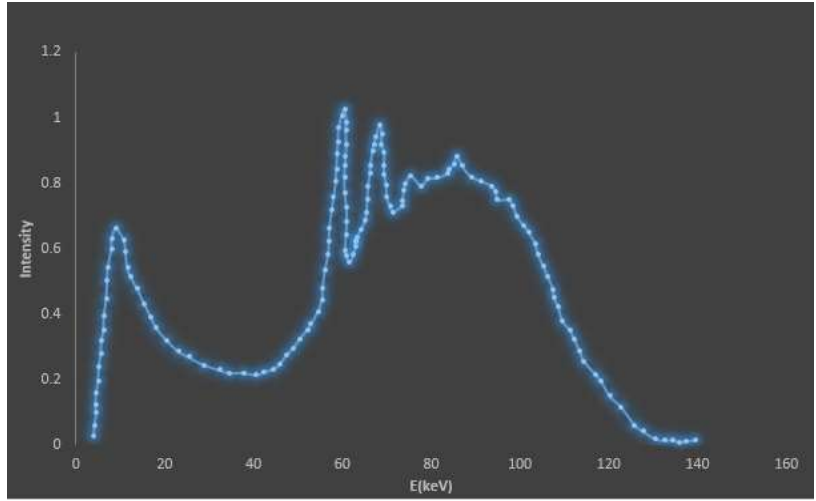

**Figure 4.** The spectrum of the x-ray tube at the energy of 140 keV recorded by the HPGe detector.

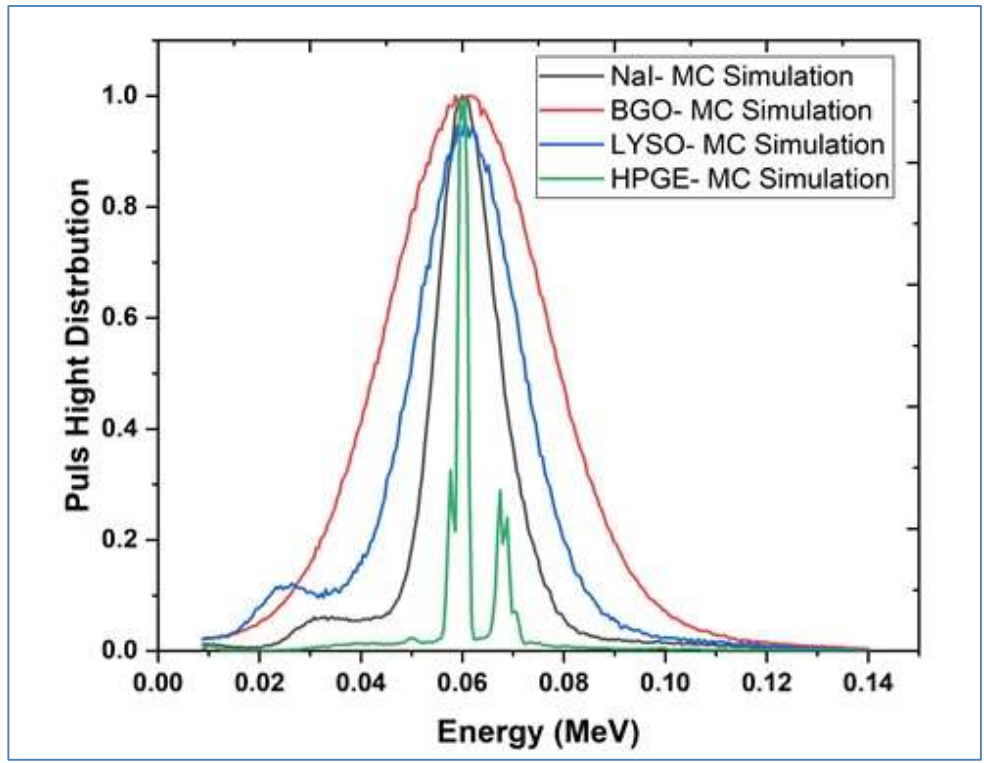

**Figure 5.** The pulse height distribution of different solid-state radiation detectors for 140 kVp x-ray radiation obtained from MC simulation.

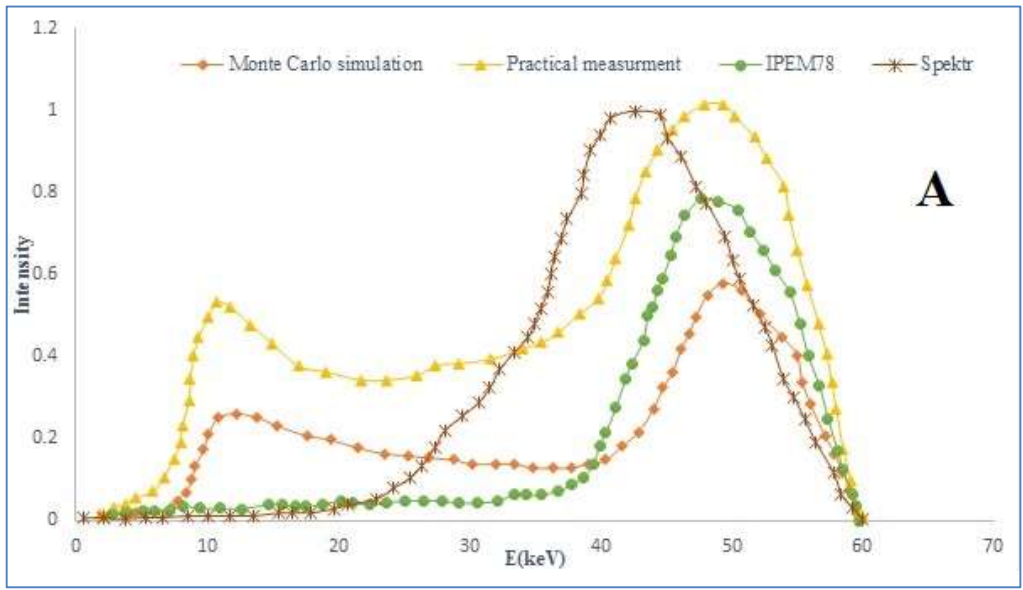

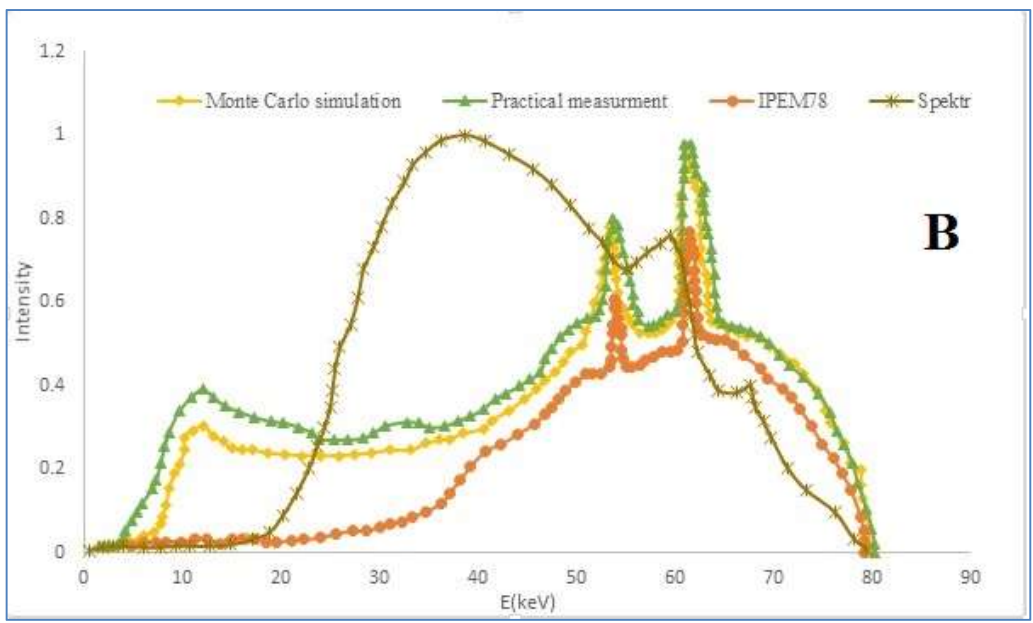

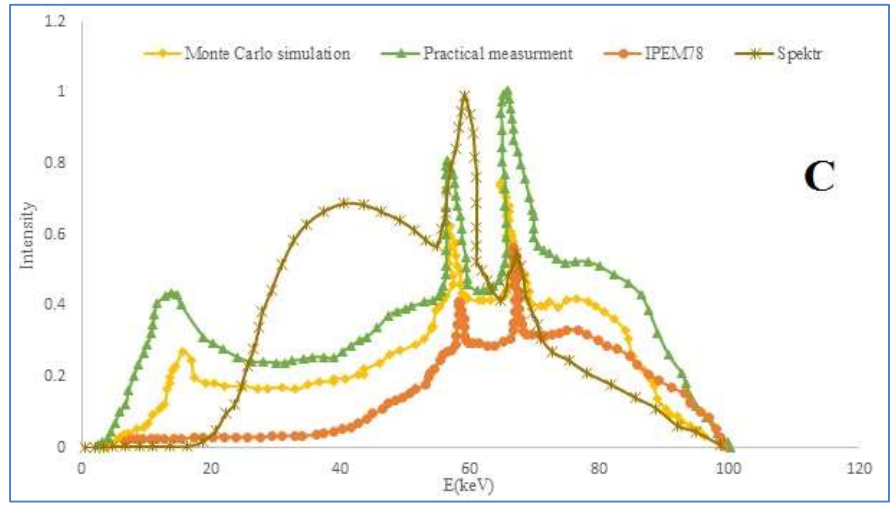
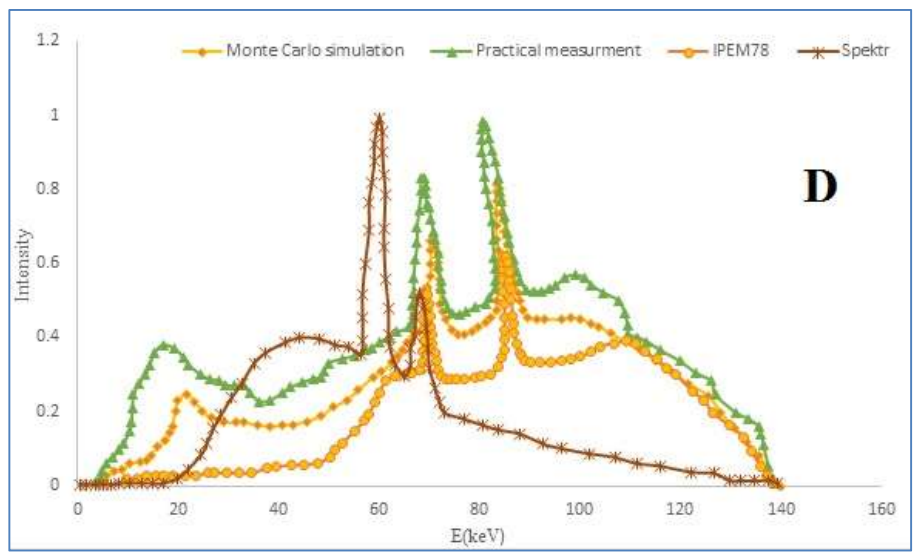

**Figure 6.** Comparison of x-spectrum estimated by MC simulation, experimental measurement, Spektr, and IPEM 78 tools for the x-ray energies of A) 60 kVp B) 80 kVp C) 100 kVp D) 140 kVp.

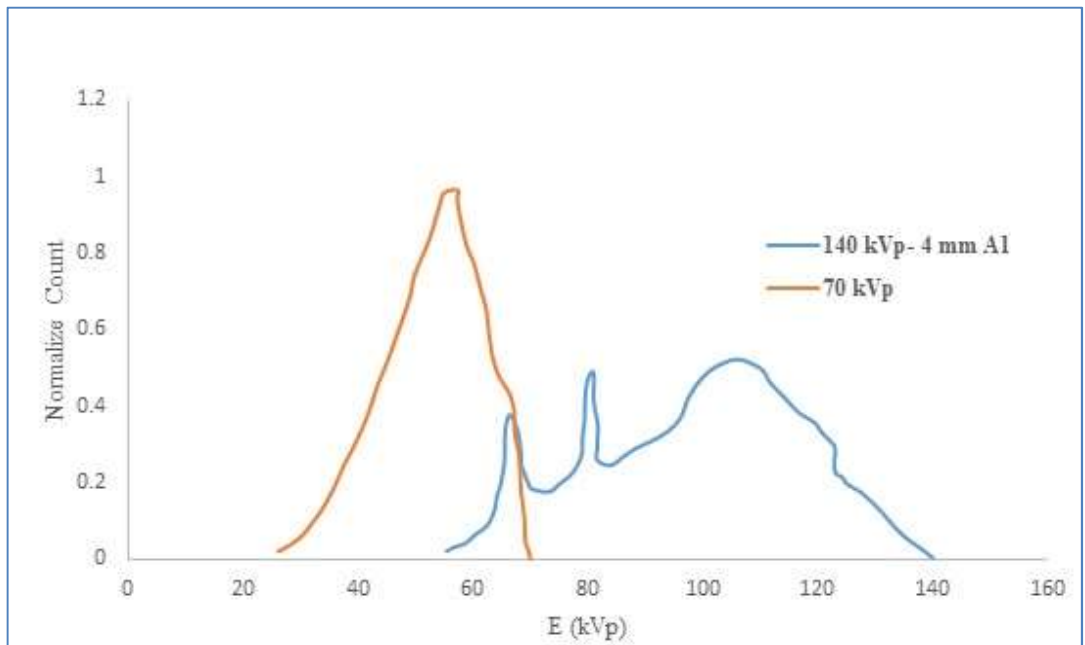

**Figure 7.** The dual-energy x-ray spectra obtained from applying 0.5 mm and 4.5 mm external filters (Al) to the original x-ray beam with the energy of 140 kVp (measured by HPGe detector).

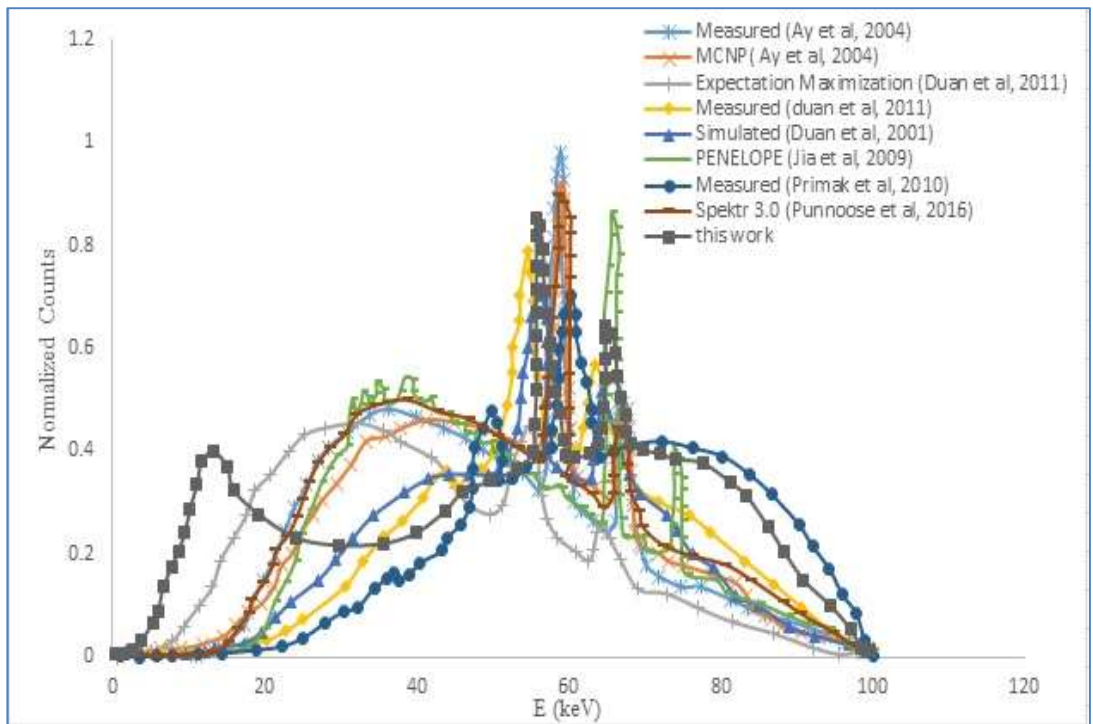

**Figure 8.** Comparison of different x-ray spectrum obtained from the different models for an x-ray beam with energy of 100 kVp.